\documentclass[12pt,titlepage,a4paper]{article}

\begin{document}
\title{Nonlinear 
three-wave interaction in marine sediments}
\author{
N.I. Pushkina\thanks
{email N.Pushkina@mererand.com}\\
M.V.Lomonosov Moscow State University, Research Computing Centre, \\ Vorobyovy Gory, 
Moscow 119992, Russia} 
\date{ }
\maketitle
\begin{abstract}
Nonlinear interaction of three acoustic waves in a sandy sediment 
is studied in the frequency range where 
there is a considerable wave velocity dispersion. The possibility of an experimental 
observation of the generation of a sound wave by two pump waves  
propagating at an angle to each other is estimated.
\end{abstract}
\section{Introduction} 

Acoustic wave propagation is an effective
tool for studying properties of porous media.
Such media are known to exhibit a considerably stronger elastic nonlinearity as compared with 
homogeneous fluid or solid media \cite{bjo, hov, bel, chot, turn}. 
This property stimulated a growing interest in studying nonlinear 
dynamics of poroelastic media (see, e.g. Refs. \cite{don, zh, daz}). 

In Refs. \cite{don, daz} 
the generation of second-harmonic acoustic waves is studied theoretically 
for the one-dimensional case. In the frequency range where there is no sound velocity dispersion, 
the energy and momentum conservation laws, 
\begin{equation}
\omega_1+\omega_2=\omega_3,\qquad{\bf k}_1+{\bf k}_2={\bf k}_3  \label{eq:sin} 
\end{equation}
are satisfied for nonlinear interactions of three waves propagating in 
one direction. Such interactions include harmonic generation of the 
fundamental wave as well as the generation of the waves with sum and difference 
frequencies of the fundamental and the excited waves. In this case, if the dissipation 
is not too
high, the formation of a shock wave can start, and a particular three wave interaction 
could be hidden.
For more explicit observation of nonlinear three wave interactions it is preferable to choose 
the frequency range where there is a significant amount of velocity dispersion. In this case the 
three waves that satisfy the conditions (\ref{eq:sin}) will propagate in three different 
directions, and no shock formation would occur. 

\section{Theory}

In the present work the nonlinear interaction of three acoustic waves propagating in a 
granual medium at an angle 
to each other is studied. It has been  experimentally observed that marine sediments
exhibit a noticeable acoustic wave velocity dispersion in some frequency 
ranges \cite{turg, sto, buch}. We shall use the dispersion velocity data \cite{turg} listed in 
Fig. 2 of Ref. \cite{buch}. The experimental data show a strong velocity dispersion within the
frequency range between 1 kHz and 10 kHz. We assume that at the boundary of the porous medium
two waves, $(\omega_1, {\bf k}_1)$ and $(\omega_2, {\bf k}_2)$, are excited at an angle 
to each other. They generate the wave $(\omega_3,{\bf k}_3)$, and let it propagate along the x-axis.
We may write 
\begin{equation}
k_1\cos\theta_1+k_2\cos\theta_2=k_3, \label{k}
\end{equation}
where the angles $\theta_1$ and $\theta_2$ are the angles between the vectors ${\bf k}_1$, 
${\bf k}_2$ and the x-axis correspondingly. The angles $\theta_1$ and $\theta_2$ depend on the 
choice of the frequencies $\omega_1$ and $\omega_2$. We shall choose them in the range of 
the maximum velocity dispersion in order the angles $\theta_1$ and $\theta_2$ are not  
too small and hence don't fall in the dissipation spreading of the waves. We shall be based 
on Fig. 2 from Ref. \cite{buch} and choose the frequencies, 
$\omega_1=2\pi\cdot2\cdot10^3s^{-1}$,  $\omega_2=2\pi\cdot3\cdot10^3s^{-1}$, for which 
the sum frequency equals\,\, 
$\omega_3=\omega_1+\omega_2=2\pi\cdot5\cdot10^3s^{-1}$. The corresponding 
acoustic wave velocities are the following:   
$c_1\approx1,58\cdot10^5cm/s,\,\, c_2\approx1,62\cdot10^5cm/s,\,\, c_3\approx1,72\cdot10^5cm/s.$
These experimental data yield the values of the three wave vectors, which allows obtaining 
the values of the angles $\theta_1$ and $\theta_2$, $\theta_1\approx25^\circ, \theta_2
\approx18^\circ.$ 

Our task is to find the amplitude of the wave $(\omega_3,{\bf k}_3)$ generated by  
the waves $(\omega_1, {\bf k}_1)$ and $(\omega_2, {\bf k}_2)$
and to estimate if the intensity of the generated wave  could reach a measurable 
value at a reasonable distance starting from the fluctuation level at the boundary. 
To solve the problem we start from the continuity equations for the densities and 
momenta of the liquid and solid phases of a sediment composed of a rigid 
frame and pores filled with water, see Refs. \cite{vn, VN}. These equations 
are equivalent in the main features to the equations developed by Biot 
\cite {Bi1, Bi2, Bi}, 
but they are presented in a different form with a more explicit 
physical meaning. On the basis of these equations, in Ref. \cite{zh} the 
equations for the densities of the liquid and solid phases, 
$\rho_f$ and $\rho_s$, were derived (in this paper we don't take 
diffraction into account):

\begin{eqnarray}
\left(1-\frac{m}{\rho_fc^2G}\right)\frac{\partial\rho_f}
{\partial\tau}-
\frac{\nu}{\rho_sc^2G}\frac{\partial\rho_s}{\partial\tau}=
\nonumber\\-c\left(1+\frac{m}
{\rho_fc^2G}\right)\frac{\partial\rho_f}{\partial x}-
\frac{\nu}{\rho_scG}\frac{\partial\rho_s}{\partial x}+\nonumber\\
\frac{1}{\rho_f}\frac{\partial\rho_f^2}{\partial\tau}+\frac{1}{c^2}
\frac{\partial P^n}{\partial\tau}; \label{eq:fl}
\end{eqnarray}
\begin{eqnarray}
\left(1-m-\frac{k+4/3\mu+\nu^2/G}{\rho_sc^2}\right)
\frac{\partial\rho_s}{\partial\tau}-\frac{m\nu}
{\rho_fc^2G}\frac{\partial\rho_f}{\partial\tau}=\nonumber\\
-c\left(1-m-\frac{k+4/3\mu+\nu^2/G}{\rho_sc^2}\right)
\frac{\partial\rho_s}{\partial x}-\frac{m\nu}
{\rho_fcG}\frac{\partial\rho_f}{\partial x}-\nonumber\\
\frac{1}{\rho_s}\left(1-m-\frac{k+4/3\mu}{\rho_sc^2}\right)
\frac{\partial\rho_s^2}{\partial\tau}
+\frac{\nu}{c^2}\frac{\partial P^n}{\partial\tau}-\frac{1}{c^2}
\frac{\partial \tilde\sigma_{xx}^n}{\partial\tau}. \label{eq:sol}
\end{eqnarray}
In these equations the following variables are used,  
\begin{equation}
x'=\epsilon x \label{eq:x}
\end{equation}
and the moving coordinate 
\begin{equation}
\tau=t-x/c. \label{eq:t}
\end{equation}
(In Eqs. (\ref{eq:fl}), (\ref{eq:sol}) and everywhere below the 
primes for $x$ are omitted.)
In the relation (\ref{eq:x}) the small parameter $\epsilon$ is 
introduced as
\begin{equation}
\epsilon \sim v_x/c \sim u_x/c \sim \delta \rho_f/\rho_f
\sim \delta \rho_s/\rho_s,
\end{equation}
here $c$ is the sound velocity in the sediment and ${v}$, ${u}$ 
are the hydrodynamic velocities of the liquid and solid phases, 
$\delta \rho_f$, $\delta \rho_s$ are the deviations from equilibrium 
values of the densities of the liquid and solid phases.
In Eqs. (\ref{eq:fl}), (\ref{eq:sol}) the left-hand sides 
are of the order of $\sim\epsilon$, the right-hand side terms are 
of the order of $\sim\epsilon^2$. 
The introduction of the new variables (\ref{eq:x}), (\ref{eq:t}) 
actually signifies the application of the method of slowly
varying wave profile at the distance of the wave-length scale.\\
In the equations (\ref{eq:fl}), (\ref{eq:sol}) we wrote $\rho_f, \rho_s$
instead of $\delta\rho_f, \delta\rho_s$, $m$ is the porosity; 
\[G=\frac{1-m}{k_s}+\frac{m}{k_f}-\frac{k}{k^{2}_{s}},\] 
where  $k_f$, $k_s$ and $k$ are the bulk moduli of the fluid, 
mineral grains constituting the frame, and of the frame itself;
$\mu$ is the shear modulus of the frame; $\nu=1-m-k/k_s$; $P^n$ is 
the nonlinear part of the pressure in the fluid, 
$\tilde\sigma_{xx}^n=\sigma_{xx}^n-k/k_sP^n$ where $\sigma_{xx}^n$ 
is the nonlinear part of the stress tensor of the 
frame. The nonlinear stress tensor is considered for 
the one-dimensional case, which corresponds to the 
accepted approximations. The explicit forms for $P^n$ 
and $\sigma_{xx}^n$ are given 
in Ref. \cite{zh} for some limiting cases. 

Let us eliminate one of the variables, $\delta \rho_f$ or 
$\delta \rho_s$, from the linear parts of Eqs. (\ref{eq:fl}), 
(\ref{eq:sol})) 
(let it be e.g. $\delta \rho_s$), by subtracting one equation 
from the other one. Note, that Eqs. (\ref{eq:fl}), (\ref{eq:sol})  
allow two independent longitudinal 
modes, the so called fast and slow waves. As it is shown in Ref. 
\cite{St}, the slow wave (unlike the fast one) is a strongly 
attenuated diffusion mode, and it does not contribute significantly 
to the sound field. In this single-mode approximation, the 
elimination of 
$\delta \rho_s$ leads to the disappearance of the linear part 
of the equation provided $c$ is the velocity of the fast wave. 
In the nonlinear parts the quantity 
$\delta \rho_s$ is expressed through $\delta \rho_f$ with the 
formula which is valid to an accuracy $\sim\epsilon$,
\begin{equation}
\delta\rho_s =\left(\frac{\nu}{\rho_sc^2G}\right)^{-1}
\left(1-\frac{m}
{\rho_fc^2G}\right)
\delta\rho_f. \label{eq:lin}
\end{equation}

 In this approximation we arrive at the 
nonlinear equation for an acoustic wave in a sediment, 
\begin{eqnarray}
\left\{2(1-m)\left[1-\left(\frac{c_f}{c}\right)^2\right]-
\frac{\nu^2}{m}
\frac{\rho_f}{\rho_s}\left(\frac{c_f}{c}\right)^2+\left
(1-m-\frac{k+4/3\mu}
{\rho_sc^2}\right)
\left[1+\left(\frac{c_f}{c}\right)^2\right]\right\}\frac
{\partial
\rho_f}{\partial x}+\nonumber\\
\frac{1}{\rho_fc}\left\{\nu\frac{\rho_f}{\rho_s}\left
[1-\left(\frac{c_f}{c}\right)^2\right]-\left[1-m-\frac{k+4/3\mu}
{\rho_sc^2}- \frac{\nu^2}{m}
\frac{\rho_f}{\rho_s}\left(\frac{c_f}{c}\right)^2\right]\right\}
\frac{\partial\rho_f^2}
{\partial\tau}-\nonumber \\
\frac{1}{c^3}\left[\left(1-m-\frac{k+4/3\mu}{\rho_sc^2}
 \right)\frac{\partial P^n}{\partial\tau}-
 \frac{\nu}{m}
\frac{\rho_f}{\rho_s}\left(\frac{c_f}{c}\right)^2\frac{\partial
\tilde\sigma_{xx}^n}{\partial\tau}\right]+a_1D_\tau\rho_f
 =0.\label{eq:D}
\end{eqnarray}
To obtain these equations we took into account that in 
sand sediments the bulk modulus $k_s$ of quartz grains is much 
greater than that of the pore water, and in this case $G$ can be 
evaluated as $G \approx m/k_f$, provided $m$ is not close to zero.  

In Eq. (\ref{eq:D}) the term $a_1D_\tau\rho_f$ that accounts for 
dissipation is introduced. $D_\tau$ is the dissipation linear operator 
in the variable $\tau$ which is characterized by the property
\begin{equation}
D_\tau e^{i\omega\tau}=\alpha(\omega)e^{i\omega\tau},\label{op} 
\end{equation}
where $\alpha$ is real and positive and it has the meaning 
of an amplitude attenuation coefficient if the coefficient 
$a_1$ is taken to be equal to the coefficient at $\partial\rho_f/\partial x$.
The relation (\ref{op}) defines the action of this operator on any 
function of the variable $\tau$ which can be represented by a 
Fourier series or integral. An algebraic expression for $\alpha(\omega)$ 
is a combination of physical parameters (complex bulk and shear frame moduli 
included) of a sediment, and it includes 
the frequency correction function introduced by Biot \cite {Bi}. 

Let us write Eq. (\ref{eq:D}) in a concise form,
\begin{equation}
a_1\frac{\partial\rho_f}{\partial x}+
a_3\frac{\partial\rho_f^2}{\partial \tau}+a_1D_\tau\rho_f=0.  \label{eq:C}
\end{equation}

Note the following. The coefficient $a_3$ at the nonlinear term in this equation 
incorporates contributions from the nonlinear stress-strain relations of 
the pore fluid, of the grains, constituting the sediment frame, and of the 
sediment frame itself. To write Eq. (\ref{eq:C}) we 
considered the nonlinear pressure $P^n$ and 
stress tensor $\tilde\sigma_{xx}^n$ from Eq. (\ref{eq:D}) as expansions in powers 
of $\delta\rho_{f,s}$,
retaining quadratic terms. The coefficient $a_3$ includes all the coefficients 
at ${\partial\rho_f^2}/{\partial \tau}$, and this coefficient will be 
estimated basing on experimental data.

We shall use Eq. (\ref{eq:C}) to find the slowly varying amplitude 
of the generated wave $\rho_3$ (from now on we shall omit the index $f$ at $\rho$).
With the three interacting waves  
written in the form $\rho_i=|\rho_i|\exp{i({\bf k}_i {\bf r}-\omega_it+\varphi_i)}$, 
$i=1,2$, and
$\rho_3=|\rho_3|\exp{i(k_3x-\omega_3t+\varphi_3)}$, and taking into account 
the relations (\ref{eq:sin}, \ref{k}), we can write Eq. (\ref{eq:C}) in
the form,
\begin{equation}
\frac{d|\rho_3|}{dx}e^{i\varphi_3}+|\rho_3|i\frac{d\varphi_3}{dx}e^{i\varphi_3}-  \label{A}
\frac{a_3}{a_1} i\omega_3|\rho_1||\rho_2|e^{i(\varphi_1+\varphi_2)}+\alpha|\rho_3|e^{i\varphi_3}=0,
\end{equation}
where $\alpha$ is the amplitude attenuation coefficient 
of the wave $\rho_3$. The second term in Eq. (\ref{A}) vanishes when the phase $\varphi_3$
attains its fixed value. The attenuation term $\alpha|\rho_3|e^{i\varphi_3}$ can be 
omitted since the wave $\rho_3$ is generated and sustained by the waves  $\rho_1$ and 
$\rho_2$ along the whole distance of their interaction. The attenuation of the 
exiting waves should be taken into account to estimate the real distance of 
the nonlinear interaction.

One can see from Eq. (\ref{A}), that the wave $\rho_3$ will "survive" and will be 
amplified in case 
$\varphi_3=\varphi_1+\varphi_2+\pi/2$, and
 the equation for the slowly varying amplitude $\rho_3$ acquires the form,
\begin{equation}
\frac{d|\rho_3|}{dx}-  \label{D}
\frac{a_3}{a_1}\omega_3|\rho_1||\rho_2|=0.
\end{equation}
Thus, the vertex (the second-order unharmonicity), that determines 
the interaction under consideration is equal to $(a_3/a_1)\omega_3$.
From this equation we get
\begin{equation}
\rho_3(l)-\rho_3(0)=  \label{l}
\frac{a_3}{a_1}\omega_3\rho_1\rho_2l, 
\end{equation}
where $\rho_3(0)$ is the amplitude (at the fluctuation level) at $x=0$, 
and it may be neglected, since $\rho_3(l)$ is supposed to be much higher.

Let us estimate the distance $l$ at which the amplitude $\rho_3$ can reach 
a measurable value. This distance cannot exceed the 
dissipation lengths of the 
waves $\rho_1$ and $\rho_2$. The amplitude attenuation coefficients of these waves 
are equal correspondingly to 
$\alpha_1\approx 0.8\cdot10^{-3}cm^{-1}$ 
and to $\alpha_2\approx3\cdot10^{-3}cm^{-1}$ (see Ref. \cite{buch}). 
This corresponds approximately to 
propagation distances $\sim1250 \,cm$ and $\sim330 \, cm$. This means, 
that $l$ cannot exceed the distance that is a little more than 
$300 \, cm$. 

The coefficient $(a_3/a_1)$ can be estimated if the second-order nonlinear parameter 
usually denoted as $B/A$ is known: $(a_3/a_1) \sim BA^{-1}/\rho c$, here $\rho$ is 
the average density of the medium. Water-saturated porous media are known to have 
a stronger nonlinearity than a homogeneous fluid. For example, the parameter $B/A$
can take values $\sim 8-12$ for sandy sediments \cite{hov, chot}, while in water it equals 
$\sim 5-6$. In Refs. \cite{z1, z2} the nonlinearity parameter of 
water-saturated sand was determined 
to be about 100. For such granual media as disordered packings of noncohesive 
elastic beads embedded in a fluid it is $10^2 - !0^3$ times larger than in 
homogeneous fluids and solids \cite{daz}. One can also note that quadratic 
nonlinearity increases significantly in the presence of gas 
bubbles in water. Experiments \cite{wu} showed that the nonlinearity parameter 
of water containing gas bubbles can reach $10^4 - 10^5$. 

We shall make a numerical estimate for the distance $l$ at which 
the ratio $\rho_3(l)/\rho_{1,2}(0)$ can reach the value 
$\sim (10^{-1}-10^{-2})$. We assume 
\begin{eqnarray*}
\rho_1(0)\sim\rho_2(0)\sim10^{-4}g/cm^3,\,\, \rho\sim 2g/cm^3,\,\, 
c=1,7\cdot 10^5cm/s,\\ 
\omega_3=2\pi\cdot5\cdot10^3s^{-1},\,\, B/A\sim 10.
\end{eqnarray*}   
These data yield the distance $l$ equal approximately to $\sim 300\, cm$, which 
falls within the attenuation length of the exciting waves.

\section{Summary}

Nonlinear interaction of three acoustic waves in a sandy sediment is studied .
A significant amount of velocity dispersion in some frequency intervals 
allows momentum and energy conservation laws to be satisfied for 
the waves propagating not in one direction. This means that no shock 
formation would occur and hence this would not obscure a three-wave nonlinear 
process. Numerical estimates 
show that the wave generated by two pump waves  
propagating at an angle to each other  in a sandy sediment can reach a 
measurable value at a 
distance realistic for an acoustic-wave experiment in a sediment.


\end{document}